# In-Field Transport Critical Currents in Superhydride Superconductors: Highly-Compressed CeH$_9$

Evgeny F. Talantsev[1,2]

[1]M.N. Miheev Institute of Metal Physics, Ural Branch, Russian Academy of Sciences, 18, S. Kovalevskoy St., Ekaterinburg, 620108, Russia

[2]NANOTECH Centre, Ural Federal University, 19 Mira St., Ekaterinburg, 620002, Russia

**Abstract**

Experimental discovery of near-room-temperature superconductivity in highly compressed hydrogen sulphate started a new era in superconductivity. To date, researchers have made the discovery of dozens of superconducting hydride phases with transition temperatures above LN2. While the primary focus of the research in this field is to determine fundamental superconducting parameters of these superconductors, here we revealed primary applied property of these superconductors which is the field dependence of transport critical current, $I_c(B,T)$. We analysed reported $V(I)$ curves for a highly compressed CeH$_9$ sample with $T_c = 70$ K, and showed that this hydride exhibits practically identical $I_c(B,T)$ to HTS 1G-wire (Bi,Pb)$_2$Sr$_2$Ca$_2$Cu$_3$O$_{11}$. Additionally, we demonstrated that the n-value in CeH$_9$ superhydride follows a nearly identical dependence on critical current, as observed in Nb$_3$Sn conductors.

**In-Field Transport Critical Currents in Superhydride Superconductors:**

**Highly-Compressed CeH$_9$**

**I. Introduction**

Experimental studies of near-room-temperature superconductors started after pivotal discovery of the superconducting state above 200 K in highly-compressed sulfur hydride by Drozdov *et al* [1] in 2015. To date, dozens of highly-compressed high-temperature superconducting hydrides have been discovered [2]–[12]. While fundamental properties/parameters of the superconducting state of the superhydrides (which are the lower critical field [1], [13], the upper critical field [14]–[17], the charge carriers density [14], [16], the London penetration depth [1], [13], [18], the coherence length [14], [16], [17], the Ginzburg-Landau parameter [13], [18]–[20], the Ginzburg-Levanyuk number [18], the energy gap [16], [21], [22], specific heat jump at the transition temperature [21], [22], the self-field critical current density [16] have been measured/extracted from experimental data), primary applied property of any superconductors, which is in-field transport critical current, $I_c(B,T)$, and its density, $J_c(B,T)$, are remaining to be practically unknown.

It should be noted that many experimental techniques utilized to measure temperature dependent superconducting gap, $\Delta(T)$, in low-temperature and high-temperature superconductors [23] are not applicable for highly compressed materials. However, the amplitude of the ground state energy gap, $\Delta(0)$, can be extracted from the self-field critical current density, $J_c(sf,T)$ [24]–[27], and the upper critical field $B_{c2}(T)$ [22], [28]. Until recent report by Semenok *et al* [15], available $I_c(B,T)$ and $J_c(B,T)$ data for superhydrides were limited by the values measured at $T \sim T_c$ [7], [9].

CeH$_9$ phase exhibits maximal transition temperature of $T_c$ = 102 K at applied pressure of $P$ = 130 GPa [15]. In this study we analyzed $V(I)$ curves measured in highly compressed CeH$_9$ sample ($P$ = 125 GPa, $T_c$ = 70 K) by Semenok *et al* [15]. It should be noted that Semenok *et al* [14] used four-probe AC method with the excitation current of 5-10 mA to measure *V*(*I*) curves, while in the applied superconductivity many research groups utilize DC transport current measurements to study HTS 2G-wires [29]. We made a comparison of the derived $I_c(B,T)$ data based on the electric field criterion of $E_c$ = 1 µV/cm with the field dependent critical current measured in well-studied superconductors.

In this study we showed that:

1. Reduced critical currents, $I_c(B,T)/I_c(B=1\ Tesla,T)$, in CeH$_9$ (*P* = 125 GPa) and 1G-wire $(Bi,Pb)_2Sr_2Ca_2Cu_3O_{11}$ are practically identical;
2. The dependence of *n*-values from critical current, $n(I_c)$, in CeH$_9$ (*P* = 125 GPa) and Nb$_3$Sn are nearly identical.

## II. Utilized model

Critical currents can be derived by the fit of measured *V*(*I*) (or *E*(*I*)) curves to the power law[30]–[32]:

$$E(I) = E_c \times \left(\frac{I}{I_c}\right)^n, \qquad (1)$$

where $E_c$ is the electric field criterion, and $n$ is the so-called *n*-value which is one of the characteristics of the transition from dissipative-less to dissipative regimes of the transport current flow in superconductors. To utilize widely used the electric field criterion value of $E_c = 1\ \mu V/cm$ [33]–[35] for CeH$_9$, the following approach was used.

Considering geometrical arrangements of experiments in the report by Semenok *et al* [15], we approximated the distance between voltage taps (electrodes) in diamond anvil cell (DAC) as $l = 50\ \mu m$. At the same time, the noise of electronics of the measured *V*(*I*) curves [15] was at the same level (< 10$^{-7}$ V) as it is in many other transport critical current measuring systems [36]–[42]. Based on that, we fitted measured *V*(*I*) curves [15] to the equation:

$$V(I) = V_0 + V_c \times \left(\frac{I}{I_{c,V}}\right)^n, \qquad (2)$$

where $V_0$ is an instrumental offset, and $V_c = 1\ \mu V$. Derived $I_{c,V}$ was recalculated to the $I_c$ by the equation:

$$I_c = \left(\frac{50\ \mu m}{1\ cm}\right)^{\frac{1}{n}} \times I_{c,V}. \qquad (3)$$

Theoretical interpretation of *n*-value in polycrystalline superconductors can be found elsewhere [30]–[32], [43].

### III. Results and Discussion

Fits of several *V(I)* curves of CeH$_9$ (*P* = 125 GPa) reported by Semenok *et al* [15] to Eq. 2 are shown in Fig. 1. We analyzed reported *V(I)* curves measured for CeH$_9$ sample with $T_c$ = 70 K. All performed fits (included fits which are not shown in Fig. 1) have a high quality (the coefficient of determination (*R*-Squared)) for each curve is better than *R* = 0.972, and for majority of the fits *R* > 0.99).

Derived $I_c(B,T=55K)$ dataset (in the reduced $I_c(B,T)/I_c(B=0.95\ Tesla,T)$ form) are shown in Fig. 2.

It should be noted that CeH$_9$ samples studied by Semenok *et al* [15] were polycrystalline samples (which can be confirmed by XRD data presented in the study [15]) with $T_c$ = 70-102 K. Based on that, it is useful to make a comparison of the deduced $I_c(B,T)/I_c(B=0.95\ Tesla,T)$ with the $I_c(B,T)/I_c(B=1.0\ Tesla,T)$ measured in HTS 1G-wires (Bi,Pb)$_2$Sr$_2$Ca$_2$Cu$_3$O$_{11}$, which are also polycrystalline samples with $T_c$ in the same ballpark ($I_c(B,T)$ data for 1G-wires reported by Wimbush and Strickland [29]). This comparison is shown in Fig. 2, where nearly identical in-field transport current dependence CeH$_9$ (*P* = 125 GPa) and two HTS 1G-wires can be seen.

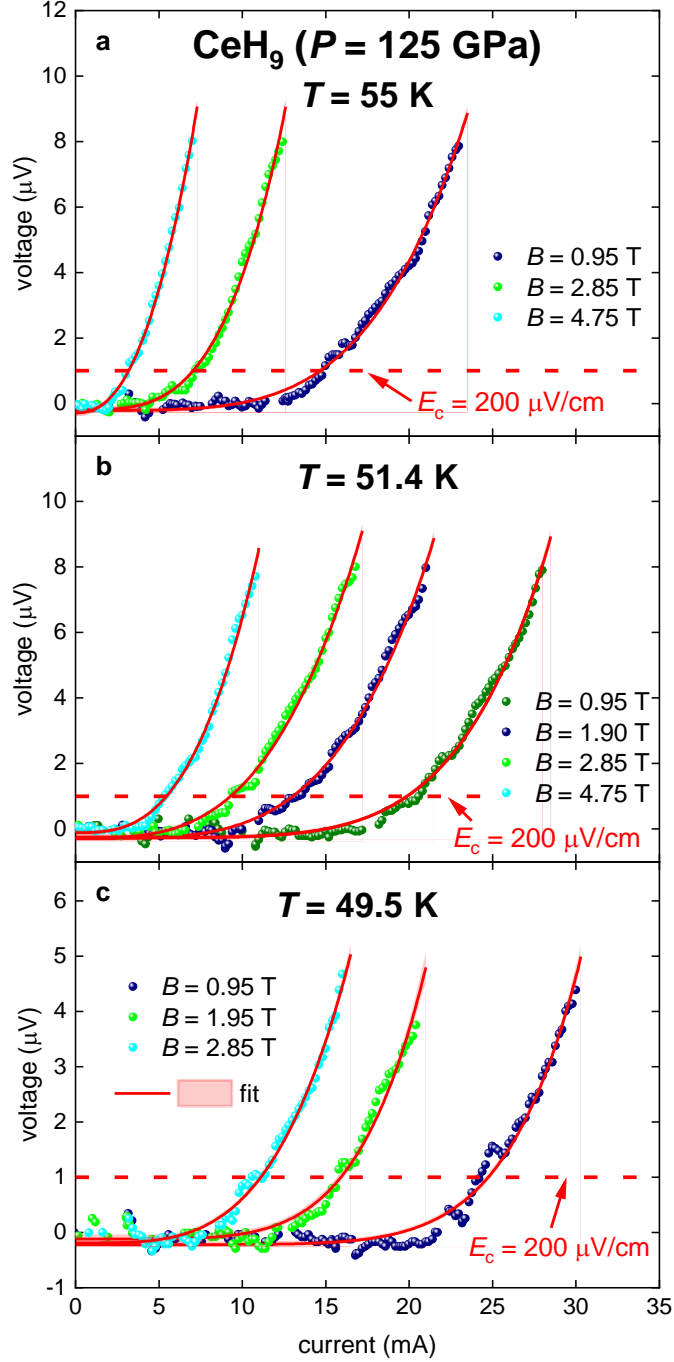

**Figure 1.** Fits of the $V(I)$ curves measured in CeH$_9$ ($P$ = 125 GPa) by Semenok *et al* [15] to Eq. 2. Raw $V(I)$ datasets are from Figure S6 [15]. Utilized voltage criterion of $V_c$ = 1 mV (or field criterion $E_c$ = 200 mV/cm) is shown. The pink shadow areas indicate 95% confidence bands for the fit to Eq. 2.

The analysis of *n*-value (Eq. 1) this is a standard practice [44]–[46] in superconductors characterization. In Fig. 3 we showed field dependence of the *n*-values for CeH$_9$ ($P$ = 125 GPa) and two HTS 1G-wires.

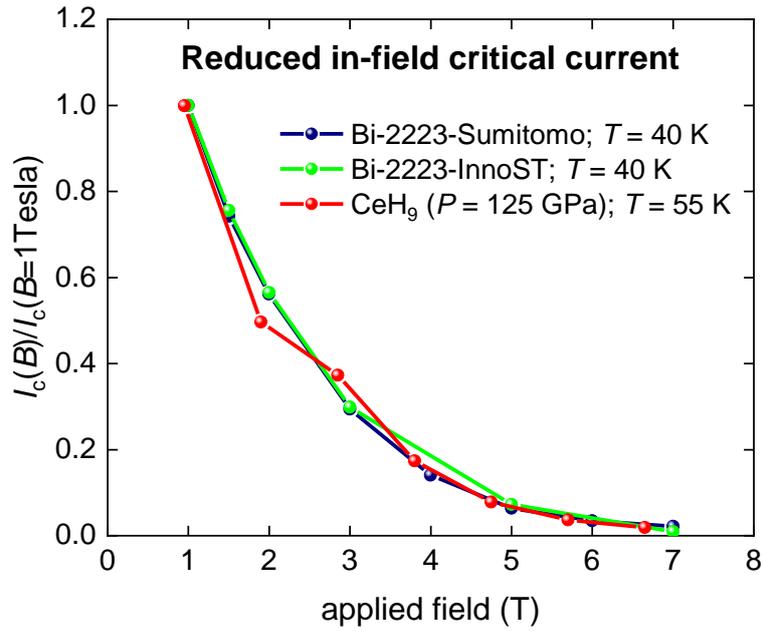

**Figure 2.** Reduced transport in-field critical currents (defined by $E_c = 1$ μV/cm criterion) for polycrystalline superconductors. Blue – Sumitomo 1G-HTS wire (data by Wimbush and Strickland [29]). Green – InnoST 1G-HTS wire (data by Wimbush and Strickland [29]). Red - CeH$_9$ ($P = 125$ GPa), raw $V(I)$ datasets reported by Semenok *et al* [15] in their Figure S6.

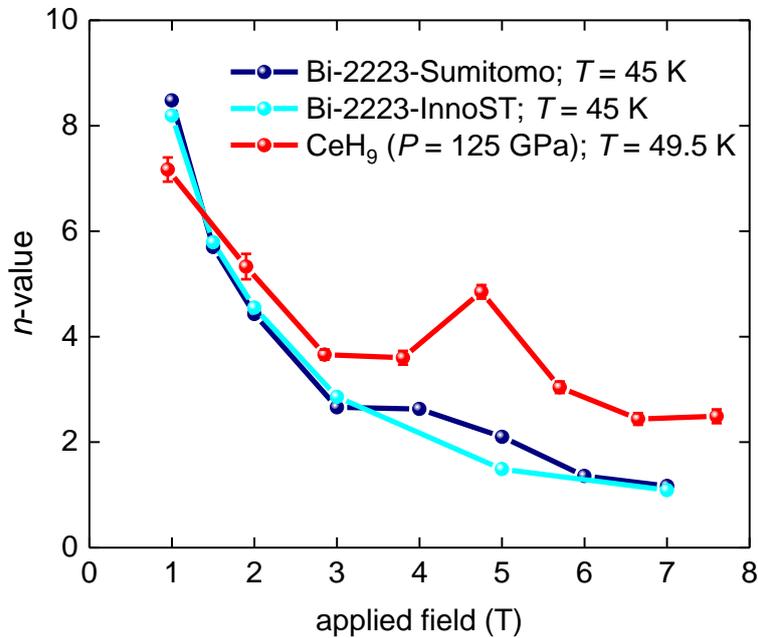

**Figure 3.** Field dependence of *N*-values for 1G-HTS wires (data by Wimbush and Strickland [29]) and CeH$_9$ ($P = 125$ GPa) (raw $V(I)$ datasets reported by Semenok *et al* [15]).

Further analysis is shown in Fig. 4, where the $n(I_c)$ dependences were fitted to the power law [45]:

$$n(I_c) = 1 + a \times (I_c)^s \qquad (4)$$

where $a$ and $s$ are free-fitting parameters. More details about parameter $s$ can be found elsewhere [47].

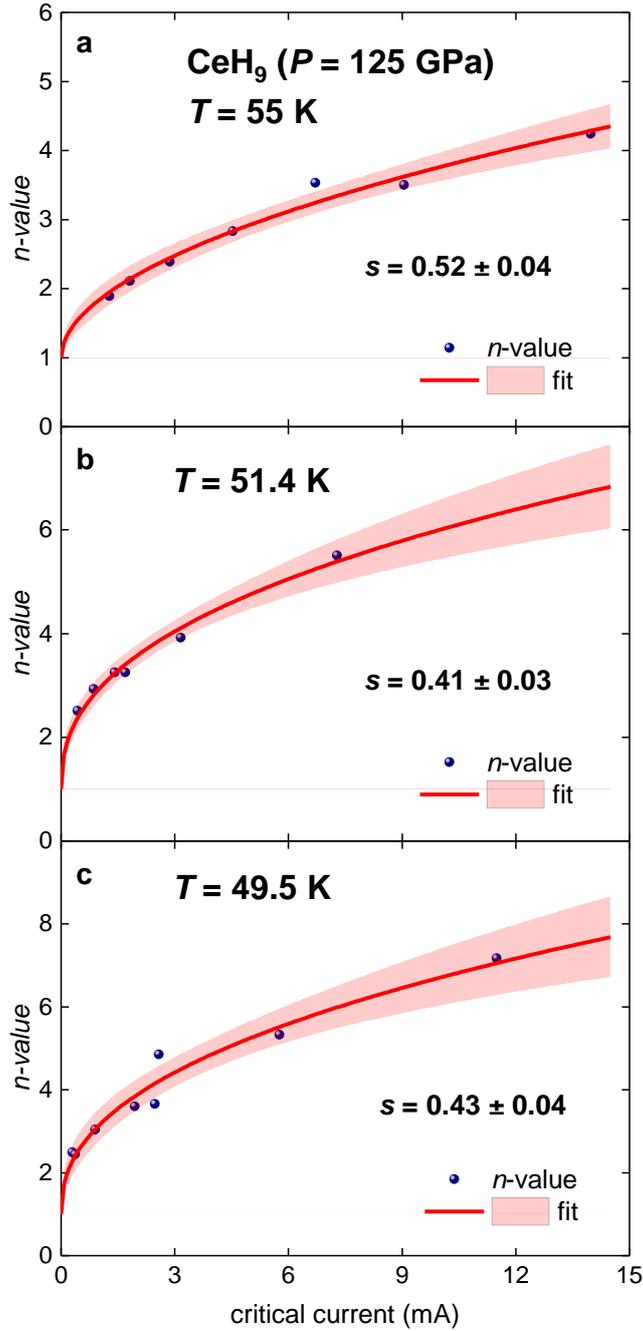

**Figure 4.** $n(I_c)$ dependences and data fits to Eq. 4 for CeH$_9$ ($P$ = 125 GPa) (raw $V(I)$ datasets reported by Semenok *et al* [15]). The pink shadow areas indicate 95% confidence bands for the fit to Eq. 4.

It is important to note that deduced $0.41 \leq s \leq 0.52$ (Fig. 4) for CeH$_9$ ($P$ = 125 GPa) are in the same ballpark values for reported by Taylor and Hampshire [45] for polycrystalline Nb$_3$Sn wires ($0.38 \leq s \leq 0.44$).

## IV. Conclusions

In this study we analyzed the only available in-field high-current $V(I)$ curves measured in highly-compressed hydrides. In the result, we derived the in-field transport critical current $I_c(B,T)$ for CeH$_9$ ($P$ = 125 GPa) hydride and showed that its field dependence is similar to one measured in polycrystalline HTS 1G-wires (Bi,Pb)$_2$Sr$_2$Ca$_2$Cu$_3$O$_{11}$. We also showed that the power-law exponent $s$ in the dependence of the $n$-value vs the critical current is close to the power-law exponent in Nb$_3$Sn.

**Acknowledgement**

The author thanks D. V. Semenok (Center for High Pressure Science and Technology Advanced Research, Beijing, PRC) and all co-workers of Ref. [15] for making their experimental data on CeH$_9$ superconductor freely available online prior the publication of their paper.

The author thanks financial support provided by the Ministry of Science and Higher Education of Russia (theme "Pressure" No. 122021000032-5). The research funding from the Ministry of Science and Higher Education of the Russian Federation (Ural Federal University Program of Development within the Priority-2030 Program) is gratefully acknowledged by the author.

**Data availability statement**

The data that support the findings of this study are available from the author upon reasonable request.

**Declaration of interests**

The author declares that he has no known competing financial interests or personal relationships that could have appeared to influence the work reported in this paper.